\documentclass[amsmath,amssymb,aps,10pt,pre,twocolumn,longbibliography]{revtex4-2}
\usepackage{graphicx}
\usepackage{bm}
\usepackage{bbm}
\usepackage{color}

\newcommand{\revis}[1]{\textcolor{black}{#1}}

\usepackage{tabularx}

\begin{document}

\title{Network analysis for steady-state current fluctuations under finite affinity: Application to Brownian computation}

\author{Yasuhiro Utsumi}
\affiliation{Department of Electrical and Electronic Engineering, Faculty of Engineering, Mie University, Tsu, 514-8507, Mie, Japan.}

\begin{abstract}
A graph-theoretic analysis of the steady-state current noise in master equations under a finite thermodynamic force (affinity) is presented. 
The incidence matrix twisted by a finite affinity is not orthogonal to the standard cycle space, motivating the introduction of twisted circuit matrices to restore the orthogonality.
The resulting twisted-cycle matrix yields an interference-like effect, enabling us to express the signal-to-noise ratio as a quadratic optimization problem in terms of twisted-cycle currents. 
We apply this framework to a Brownian computation model on a tree-like state-transition diagram with exponential backward branching, finite affinity at each step, and a single reset cycle. 
In the limit of an infinitely long intended computation path $\ell$, the Fano factor of the reset current undergoes a transition from noiseless to Poissonian behavior at an affinity equal to the logarithm of the number of immediate predecessors $\alpha$. 
This corresponds to an easy-hard transition in the computational time complexity [K.~Okajima, K.~Hukushima, arXiv:2512.24728
], which is not captured by the thermodynamic uncertainty relation. 
This transition point precisely characterizes the thermodynamic costs of logically irreversible computation: 
in the absence of affinity, the reset cost scales as $\ln \ell$, whereas reaching the transition point requires a thermodynamic force of order $\ln \alpha$ per step to counteract backward branching. 
\end{abstract}

\date{\today}
\maketitle

\newcommand{\mat}[1]{\mbox{\boldmath$#1$}}

\section{Introduction}

Graph theory has been a standard tool for analyzing electric circuit networks~\cite{Bryant1961,Branin1967,Takahashi1969} and also for superconducting quantum circuit networks~\cite{RasmussenPRX2021}. 
For nonequilibrium systems described by Markov master equations, graph-theoretic approaches have also proven powerful, since the stochastic dynamics is influenced by the topology of the underlying state-transition diagram~\cite{Schnakenberg1976}; these approaches have been further developed in the context of chemical reaction networks, e.g., ~ \cite{DalCengio2023,Yoshimura2023}. 
Graph-theoretic approaches have been applied to analyze fluctuation relations~\cite{Andrieux2007,Polettini2014} and thermodynamic uncertainty relations (TURs)~\cite{Pietzonka2016,Polettini2016,Utsumi2024}.

The Brownian computer~\cite{Bennett1982} is a prototypical model in the thermodynamics of computation, 
a field that has recently attracted renewed attention~\cite{Wolpert2024,Manzano2024}. 
Motivated by models of Brownian computation with resets~\cite{Utsumi2022computation,Utsumi2023}, whose state-transition diagrams form large graphs with a small number of cycles, we previously introduced a transformation that expresses current fluctuations in terms of cycle currents~\cite{Utsumi2024}. 
The idea behind this transformation is reminiscent of the vortex--charge duality~\cite{Nagaosa1999} in Josephson junction arrays~\cite{Fazio1991} and superconducting circuits~\cite{Ulrich2016}.

In the previous analysis~\cite{Utsumi2024}, we limited ourselves to uniform bidirectional and unidirectional transition rates. 
In the present paper, we extend it to arbitrary transition rates. 
\revis{
The conceptual novelty at finite thermodynamic force or affinity, which plays the role of a gauge potential~\cite{Polettini2012}, is the coexistence of the standard incidence matrix entering the continuity equation (\ref{eqn:master_eq}) and a twisted incidence matrix arising in the finite-affinity gradient-flow (\ref{eqn:j_D_twist}). 
At zero affinity, these two matrices coincide, whereas at finite affinity the cut space associated with the twisted incidence matrix is not orthogonal to the cycle space defined in the standard manner.
}
To address this issue, 
\revis{
we construct twisted cut and cycle spaces and modify the corresponding fundamental circuit matrices so as to restore their orthogonality in Sec.~\ref{sec:Twisted_circuit_matrices}. 
This constitutes the genuinely new mathematical ingredient of the present work and allows the quadratic optimization structure in cycle-current space developed in Ref.~\cite{Utsumi2024} to be directly generalized to finite affinity in Sec.~\ref{sec:Duality_transformation}. 
The resulting twisted circuit matrices induce interference-like effects associated with finite cycle affinities. 
}
To our knowledge, such a seemingly straightforward decomposition has not been explicitly discussed in the context of current-noise analysis.

We apply our formalism to a logically irreversible Brownian computation process on an exponentially backward-branched $\alpha$-ary tree-like graph with a single reset cycle. 
In this model, the state-transition diagram is almost acyclic, except for the reset cycle. 
We show that the Fano factor changes from 0 to 1, while the reciprocal of the average reset current, corresponding to the computation time, changes from linear to exponential scaling with the length of the intended computation path. 
This transition point corresponds to the easy--hard transition observed in numerical simulations of first-passage time without reset~\cite{Okajima2026}.
We further discuss the information thermodynamic cost within the framework of stochastic thermodynamics of resetting~\cite{Fuchs2016,Utsumi2022computation,Utsumi2023}. 
This naturally associates the reset cost~\cite{Norton2013,Strasberg2015,Utsumi2023} with the zero-affinity limit, and the driving force $k_{\rm B} T \ln \alpha$ at temperature $T$~\cite{Bennett1982} with the easy--hard transition point~\cite{Okajima2026}.

The structure of the paper is as follows.
In Sec.~\ref{sec:snr}, we summarize the signal-to-noise ratio as a quadratic optimization problem, which serves as the starting point of our analysis.
In Sec.~\ref{sec:circuit_matrices}, we review the circuit matrices and introduce their twisted counterparts.
We then perform a duality transformation to express the signal-to-noise ratio in terms of the cycle currents.
In Sec.~\ref{sec:examples}, we present two examples.
The first example is a graph containing two cycles, which highlights interference-like effects arising from finite cycle affinities.
The second example is a schematic model of Brownian computation, which includes a discussion of the information-thermodynamic cost.
Section~\ref{sec:conclusion} summarizes our results.
Throughout the paper, we use the notations and technical terms adopted in Ref.~\cite{Utsumi2024}.

\section{quadratic optimization formulation of current noise}
\label{sec:snr}

The state transition diagram of a continuous-time Markov chain is a directed multigraph $G_{\rm m}=(V,E_{\rm m})$, 
where the set of nodes $V$ and the set of arcs (directed edges) $E_{\rm m}$ represent states and transitions, respectively. 
We focus on a connected graph with a unique steady state for simplicity. 
The direction of an arc corresponds to the direction of a transition. 
The set of arcs can be partitioned into mutually disjoint sets of arcs for forward transitions $E$ and backward transitions $\overline{E}$ as, 
$E_{\rm m} = E \sqcup \overline{E}$. 
Here and hereafter, we use the overline to represent the set of reversed arcs, 
$\overline{A} = \left \{ -e \middle | e \in A \right \}$. 
We assume the transition rate associated to the forward process $e \in E$ is always positive $\Gamma_e > 0$, 
while that associated with the backward process can be zero, $\Gamma_{-e}=0$. 
Therefore, 
$E=E_{\rm b} \sqcup E_{\rm uni}$, where 
$E_{\rm uni} = \left \{ e \middle | e \in E \wedge (\Gamma_{-e}=0) \right \}$ 
is for the unidirectional processes and 
$E_{\rm b} = \left \{ e \middle | e \in E \wedge (\Gamma_{-e}>0) \right \}$ 
for the bidirectional processes.  

In the following, we focus on the case where there exists a directed graph $G=(V, E)$ that contains a directed rooted spanning tree $T=(V(T), E(T))$, with $V(T)=V$, in which each arc is oriented away from the root node $v_0 \in V$. 
We assume that every twig, i.e., an arc of the spanning tree $t \in E(T)$, corresponds to a bidirectional process $t \in E_{\rm b}$. 
We refer to $T^*$ as the cotree of $T$, which consists of arcs not in $T$, so that $E(T) \sqcup E(T^*) = E$. 
Each chord $c \in E(T^*)$ completes a unique fundamental cycle. 
It can correspond to either a bidirectional or a unidirectional process $\Gamma_{-c} \geq 0$.

We write the arc $e \in E_{\rm m}$ from the tail node $\partial^+ e \in V$ to the head node $\partial^- e \in V$ as a tuple $e=(\partial^- e \leftarrow \partial^+ e)$, 
where $\partial^\pm$  stands for the boundary operator. 
The positive (negative) incidence matrix is defined as,  
$D^{\pm}_{v,e} = \delta_{v,\partial^\pm e}$. 
The reversed arc of $e$ is then $-e=(\partial^+ e \leftarrow \partial^- e)$, which satisfies, 
$D^{\pm}_{v,-e} = D^{\mp}_{v,e}$. 
The incidence matrix ${\bm D} = {\bm D}^+ - {\bm D}^- \in \mathbb{R}^{|V| \times |E|}$, works as the divergence operator in the vector analysis. 
Hereafter, the number of elements of a set $A$ is denoted as $|A|$. 
The master equation is written in the form of the continuity equation as, 
\begin{align}
\dot{ {\bm n} } = - {\bm D} {\bm j}({\bm n}) \, , \label{eqn:master_eq}
\end{align}
where ${\bm n} = \begin{bmatrix} n_{v_1},n_{v_2}, \dots, n_{v_{|V|}} \end{bmatrix}^{\sf{T}} \in \mathbb{R}^{|V|}$ ($v_1, v_2, \dots, v_{|V|} \in V$) is a vector of node state probabilities. 
${\bm j} \in \mathbb{R}^{|E|}$ is a vector of edge currents, 
\begin{align}
j_e(n_v) = \sum_{v \in V} \left ( \Gamma_{e} D_{v,e}^+ - \Gamma_{-e} D_{v,e}^- \right ) n_v \, . \label{eqn:edge_current}
\end{align}

In the present paper, we will focus on the fluctuations of the weighted sum of edge currents, 
\begin{align}
w = \sum_{e \in E} d_e j_e \, , \;\;\;\; ( d_e \in \mathbb{R} ) \, . \label{eqn:d} 
\end{align}
The average value in the steady-state reads, 
$\langle \! \langle w \rangle \! \rangle = {\bm d}^{\sf{T}} {\bm j} \left ( {\bm n}^{\rm st} \right )$, 
where ${\bm d} \in \mathbb{R}^{|E|}$. 
The steady-state node state probability ${n}_v^{\rm st}$ satisfies $\dot{n}_v^{\rm st}=0$ and can be calculated using Kirchhoff-Hill theorem~\cite{Schnakenberg1976,Weidlich1978}. 

Within the Gaussian approximation, the probability distribution of $w$ in the limit of long measurement time $\tau$, reads~\cite{Utsumi2024}, 
\begin{align}
\frac{\ln {\mathcal P}_\tau ( w )}{\tau} \approx - \frac{\left ( w - \langle \! \langle w \rangle \! \rangle \right )^2 }{ \revis{2} \langle \! \langle w^2 \rangle \! \rangle} \, , 
\end{align}
where $\langle \! \langle w^2 \rangle \! \rangle$ represents the scaled second cumulant determined by solving the quadratic optimization problem:
\begin{align}
\frac{\langle \! \langle w \rangle \! \rangle^2}{\langle \! \langle w^2 \rangle \! \rangle} = \inf_{ {\bm j}^\perp \in J^{(1)}, \, {\bm \phi} \in P^{(1)} } 
\left( {\bm j}^{\perp} - {\bm j}\left( {\bm \phi} \right) \right)^{\sf{T}} {\bm G}^{-1} 
\left( {\bm j}^{\perp} - {\bm j}\left( {\bm \phi} \right) \right) \, , \label{eqn:Sigma_w}
\end{align}
where 
${\bm G} ={\rm diag} \,{\bm g}({\bm n}^{\rm st})$ with $g_e(n_v) = a_e(n_v) + a_{-e}(n_v)$ is the inverse of diagonal weight matrix. 
The constraints are, 
\begin{align}
J^{(1)} =& \left \{ {\bm j}^\perp \in {\mathbb R}^{|E|} \middle |  \left( \langle \! \langle w \rangle \! \rangle = {\bm d}^{\sf{T}} {\bm j}^{\perp}  \right) \wedge 
\left(  {\bm j}^{\perp} \in \ker {\bm D} \right) \right \}  \, , \label{eqn:set_jw1}
\\
P^{(1)}=& \left \{ {\bm \phi} \in {\mathbb R}^{|V|} \middle | {\bm 1}^{\sf{T}} {\bm \phi} = 0 \right \}  \, , \label{eqn:set_p1}
\end{align}
where ${\bm 1}$ is a real vector whose entries are ones. 
Equation (\ref{eqn:Sigma_w}) (for the derivation, see, e.g., Ref.~\cite{Utsumi2024}) is the starting point of the following analysis.

\section{Twisted circuit matrices and Duality transformation}
\label{sec:circuit_matrices}

In the following, we fix the ordering of the elements of the edge current vector by listing first twig currents and then chord currents: 
${\bm j}^{\sf{T}} = \begin{bmatrix} {\bm j}_t^{\sf{T}} & {\bm j}_{c}^{\sf{T}} \end{bmatrix} \in \mathbb{R}^{|E|}$, 
where 
${\bm j}_t^{\sf{T}} = \begin{bmatrix} j_{t_1}, j_{t_2}, \dots, j_{t_{|E(T)|}} \end{bmatrix}$
and 
${\bm j}_c^{\sf{T}} = \begin{bmatrix} j_{c_1}, j_{c_2}, \dots, j_{c_{|E(T^*)|}} \end{bmatrix}$. 
In this section, we use letters $e$, $t$, and $c$ to express an arc $e \in E$, a twig $t \in E(T)$, and a chord $c \in E(T^*)$.  

\subsection{Circuit matrices}

We summarize standard circuit matrices (see, e.g., Refs.~\cite{Bryant1961, Branin1967,Takahashi1969,RasmussenPRX2021,Utsumi2024}). 
Along the directed rooted spanning tree $T$, there is a unique directed path from the root $v_0$ to a node $v_\ell$. 
We write this path as a sequence of twigs $t_1,t_2,\dots,t_\ell \in E(T)$ and nodes $v_0,v_1,\dots,v_\ell \in V(T)$. 
In right-to-left order, it is written as
\begin{align}
P_{v_\ell \leftarrow v_0} =(v_\ell,t_\ell, v_{\ell-1}, \dots , t_2,v_1,t_1,v_0) \, .
\label{eqn:path}
\end{align} 
The number of arcs $\ell$ corresponds to the length of the path. 
The head of $t_n$ and tail of $t_{n+1}$ ($n=1,\dots,\ell-1$) share the same node $v_n$, i.e., $v_n = \partial^- t_{n} = \partial^+ t_{n+1}$. 
The two endpoints are $\partial^+ t_1 = v_0$ and $\partial^- t_\ell = v_\ell$. 
In the following, we sometimes omit the nodes in the path. 

The root-to-node path matrix ${\bm S} \in \mathbb{R}^{|V| \times |E(T)|}$~\cite{Bryant1961,Branin1967,Takahashi1969,Utsumi2024}, is defined in the way that its $(v,t)$ element is 1(0) if a twig $t$ is in (not in) the directed path from the root $v_0$ to the node $v$: 
\begin{align}
S_{v,t} = \mathbbm{1}_{P_{v \leftarrow v_0}}(t) = \mathbbm{1}_{V(T_{\partial^- t})}(v) \, , 
\label{eqn:pathntegratormatrix}
\end{align} 
where the indicator function $\mathbbm{1}_{A}(a)$ equals $1$ if $a \in A$ and equals $0$ if $a \notin A$. 
Here, $T_{\partial^- t}$ is a subtree rooted at $\partial^- t \in V$ obtained by cutting the directed rooted spanning tree $T$ by removing the twig $t$. 
The right-hand side of (\ref{eqn:pathntegratormatrix}) implies another interpretation that $S_{v,t}$ is 1(0) if a node $v \in V$ is in (not in) the subtree $T_{\partial^- t}$. 

The fundamental cutset matrix ${\bm Q} \in \mathbb{R}^{|E(T)| \times |E|}$ is introduced as, 
\begin{align}
{\bm Q}^{\sf{T}} = - {\bm D}^{\sf{T}} {\bm S} =   \begin{bmatrix} {\bm I}_{|E(T)|} & {\bm F} \end{bmatrix}^{\sf{T}} \, , \label{eqn:cutsetmatrix}
\end{align} 
where ${\bm I}_{|E(T)|}$  is a $|E(T)| \times |E(T)|$ unit matrix. 
The $(t,c)$ element of ${\bm F} \in \mathbb{R}^{|E(T)| \times |E(T^*)|}$
is given by, 
\begin{align}
F_{t,c} = \mathbbm{1}_{C_c}(-t) - \mathbbm{1}_{C_c}(t) \, . 
\label{eqn:F_matrix}
\end{align} 
Here, $C_c$ is the fundamental cycle associated with the chord $c$. 
It consists of the chord $c \in E(T^*)$ and the unique path in the spanning tree connecting its endpoints. 
Assuming the cycle length is $\ell$, it is written as
\begin{align}
C_c = (c,e_{\ell-1}, \dots, e_1) \, ,
\end{align}
where $e_1, \dots, e_{\ell-1} \in E(T) \cup \overline{E(T)}$. 
The last and the first arcs satisfy $\partial^- c=\partial^+ e_1$, so that the sequence forms a closed cycle with orientation determined by $c$. 
Along this ordering (from right to left), the sequence of twigs consists of those first traversed opposite to their reference orientation, followed by those traversed along it. 

Intuitively, the transpose of the incidence matrix, ${\bm D}^{\sf T}$, plays the role of a gradient operator in vector analysis.
Therefore, the relation $Q_{t,t'} = \delta_{t,t'}$ in (\ref{eqn:cutsetmatrix}), where $t,t' \in E(T)$, implies that $S_{v,t}$ means the line integral along the directed rooted spanning tree from the head node of $t$ to the node $v$.
A fundamental cutset vector, the $t$-th row of ${\bm Q}$, 
corresponds to the set of arcs crossing the boundary of the subtree $T_{\partial^- t}$: 
its $e$ component, $Q_{t,e}$, is $1$ ($-1$) when the arc $e$ enters (leaves) the subtree $T_{\partial^- t}$, 
see Fig.~\ref{fig:graph_interference}.

The fundamental cycle matrix ${\bm B} \in \mathbb{R}^{|E(T^*)| \times |E|}$ 
indicates the participation and orientation of each arc in the $|E(T^*)|$ fundamental cycles: 
\begin{align}
{\bm B} = \begin{bmatrix} -{\bm F}^{\sf{T}} & {\bm I}_{|E(T^*)|} \end{bmatrix} \, . \label{eqn:cyclematrix}
\end{align} 
The fundamental cutset and cycle matrices satisfy, 
\begin{align}
{\bm Q} {\bm B}^{\sf{T}} = {\bm 0}_{|E(T)| \times |E(T^*)|} \,, \label{eqn:Tellegen_bare}
\end{align} 
which expresses that the cycle space is orthogonal to the cut space, a relation underlying Tellegen's theorem in electric circuit networks~\cite{GrossBook2004}. 
From (\ref{eqn:cutsetmatrix}) and (\ref{eqn:cyclematrix}), $\operatorname{rank} {\bm Q}=|E(T)|$ and $\operatorname{rank} {\bm B}=|E(T^*)|$. 
Since $\dim \, \operatorname{im} {\bm B}^{\sf{T}} = \operatorname{rank} {\bm B}$ and $\dim \, \ker {\bm Q} = |E| - \operatorname{rank} {\bm Q} =|E(T^*)|$, it follows that 
$\dim  \operatorname{im} {\bm B}^{\sf{T}} = \dim \ker {\bm Q}$. 
Since (\ref{eqn:Tellegen_bare}) implies $\operatorname{im} {\bm B}^{\sf{T}} \subseteq \ker {\bm Q}$, we obtain 
\begin{align}
\operatorname{im} {\bm B}^{\sf{T}} = \ker {\bm Q} \, . \label{eqn:imBTequivKerQ}
\end{align} 
Therefore, the edge current space admits the decomposition 
$ \mathbb{R}^{|E|} = \operatorname{im} {\bm Q}^{\sf{T}} \oplus \operatorname{ker} {\bm Q} = \operatorname{im} {\bm Q}^{\sf{T}} \oplus \operatorname{im} {\bm B}^{\sf{T}}$. 

The incidence matrix and the cycle matrix also satisfy, ${\bm D} {\bm B}^{\sf{T}} = {\bm 0}_{|V| \times |E(T^*)|}$, which corresponds to $\mathrm{div}(\mathrm{curl}) = 0$ in vector analysis~\cite{GrossBook2004}. 
For the connected graph, we obtain~\cite{Utsumi2024}, 
\begin{align}
\mathrm{im} {\bm B}^{\sf{T}} = \ker {\bm D} \, . \label{eqn:imBTequivKerD} 
\end{align} 

\subsection{Twisted circuit matrices}
\label{sec:Twisted_circuit_matrices}

We extend the previous discussion to the incidence matrix twisted by a finite thermodynamic force (affinity),
\begin{align}
A_e = \ln \frac{\Gamma_e}{\Gamma_{-e}} \, .
\label{eqn:affinity}
\end{align}
For unidirectional processes $e \in E_{\rm uni}$, we introduce an auxiliary backward transition rate $\Gamma_{-e}=\eta$ and take the limit $\eta \to 0$ after completing the calculations.

We define the twisted incidence matrix as
\begin{align}
\tilde{D}_{v,e} = \Gamma_e D_{v,e}^+ - \Gamma_{-e} D_{v,e}^- \, ,
\label{eqn:D_twist}
\end{align}
in terms of which the edge current vector (\ref{eqn:edge_current}) 
\revis{takes the gradient-flow form},  
\begin{align}
{\bm j}({\bm n}) = \tilde{\bm D}^{\sf{T}} {\bm n} \, .
\label{eqn:j_D_twist}
\end{align}

The cut space associated with this twisted incidence matrix is no longer orthogonal to the standard cycle space, i.e., $\tilde{\bm D} {\bm B}^{\sf{T}} \neq {\bm 0}_{|V| \times |E(T^*)|}$ in general. 
To restore the relation (\ref{eqn:Tellegen_bare}), we redefine the circuit matrices. 
Motivated by a Peierls substitution, we extend the root-to-node path matrix (\ref{eqn:pathntegratormatrix}) as
\begin{align}
\tilde{S}_{v,t} = \mathbbm{1}_{P_{v \leftarrow v_0}}(t) \frac{ e^{A_{P_{v \leftarrow \partial^- t}}} }{\Gamma_{-t}} \, ,
\label{eqn:gpathmat}
\end{align}
where $A_P = \sum_{e \in P} A_e$ denotes the path affinity, representing the accumulation of the thermodynamic force along the directed path $P$. 
Noting that each twig always corresponds to a bidirectional process, so that $e^{\pm A_{t'}}$ for $t' \in E(T)$ is finite, the twisted fundamental cutset matrix defined in the same way as (\ref{eqn:cutsetmatrix}) becomes (Appendix~\ref{app:modified_cutset_matrix}), 
\begin{align}
\tilde{\bm Q} = - \tilde{\bm S}^{\sf{T}}  \tilde{\bm D} = \begin{bmatrix} {\bm I}_{|E(T)|} & \tilde{\bm F} \end{bmatrix} \, , \label{eqn:Q_twist}
\end{align} 
which leads to the definition of the twisted fundamental cycle matrix,
\begin{align}
\tilde{\bm B} = \begin{bmatrix} -\tilde{\bm F}^{\sf{T}} & {\bm I}_{|E(T^*)|} \end{bmatrix} \, , \label{eqn:B_twist}
\end{align} 
that preserves the orthogonality between the twisted cycle and cut spaces, 
\begin{align}
\tilde{\bm Q}  \tilde{\bm B}^{\sf{T}} = {\bm 0}_{|E(T)| \times |E(T^*)|} \, .  \label{eqn:Tellegen_t}
\end{align} 
Then, by repeating the same argument used to derive (\ref{eqn:imBTequivKerQ}) from (\ref{eqn:Tellegen_bare}), we obtain
\begin{align}
\operatorname{im} \tilde{\bm B}^{\sf{T}} = \ker \tilde{\bm Q} \, . \label{eqn:imtBTequivKertQ}
\end{align} 

\begin{figure}[ht]
\begin{center}
\includegraphics[width=1 \columnwidth]{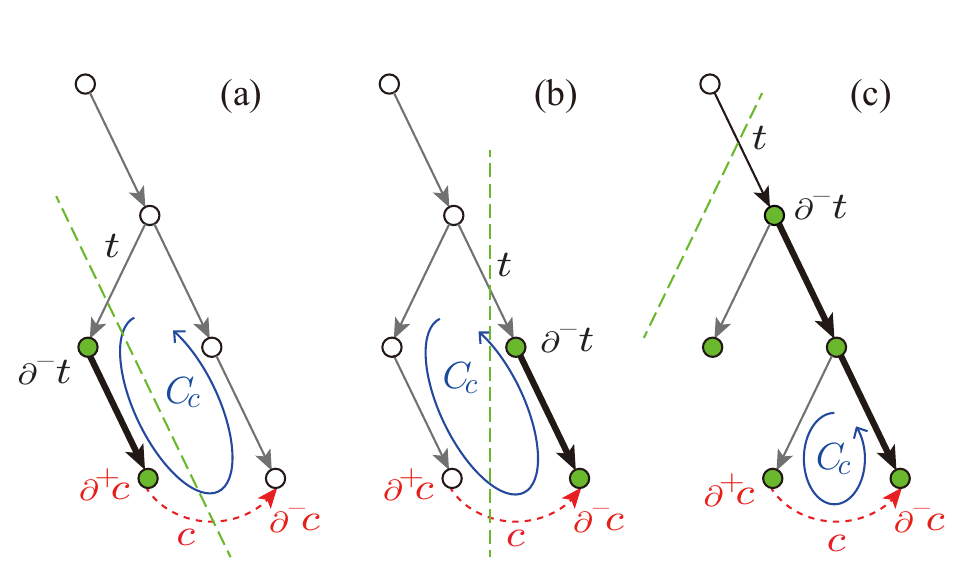}
\caption{
Three typical configurations of the twig $t$ and the chord $c$ that yield nonvanishing $\tilde{F}_{t,c}$. 
(a) $t \in C_c$, (b) $-t \in C_c$ and (c) $\partial^\pm c \in V(T_{\partial^- t})$. 
In panel (a), a thick arc indicates the path $P_{\partial^+ c \leftarrow \partial^- t}$, 
while in panels (b) and (c), thick arcs indicate the path $P_{\partial^- c \leftarrow \partial^- t}$. 
In each panel, a dotted arrow indicates the chord $c$. 
The nodes of the subtree $V(T_{\partial^- t})$ are indicated by filled circles.
The boundary between $V(T_{\partial^- t})$ and its complement is indicated by a dashed line. 
The dashed line intersects the arcs of the corresponding fundamental cutset.
}
\label{fig:graph_interference}
\end{center}
\end{figure}

The matrix $\tilde{\bm F}$ introduced above is modified by the path affinity. 
Panels (a), (b), and (c) of Fig.~\ref{fig:graph_interference} show three typical configurations of the twig $t$ and the chord $c$ that yield nonvanishing $\tilde{F}_{t,c}$. 
When the twig $t$ lies on the cycle $C_c$ with the same orientation, $t \in C_c$ [panel (a)],
\begin{align}
\tilde{F}_{t,c} = - \frac{\Gamma_{-c}}{\Gamma_{-t}} e^{A_c+A_{P_{\partial^+ c \leftarrow \partial^- t}}} 
\, . \label{eqn:Tt_para}
\end{align}
When the twig $t$ lies on the cycle $C_c$ with the opposite orientation, $-t \in C_c$ [panel (b)],
\begin{align}
\tilde{F}_{t,c} = \frac{\Gamma_{-c}}{\Gamma_{-t}} e^{A_{P_{\partial^- c \leftarrow \partial^- t}}} \, .  \label{eqn:Tt_antipara}
\end{align}
These expressions reduce to (\ref{eqn:F_matrix}) for uniform transition rates where any nonvanishing transition rate is unity and thus $A_e=0$ ($e \in E_{\rm b}$). 
In panel (c), the cycle $C_c$ does not contain the twig $t$, but is located downstream of the twig $t$ along the spanning tree $T$, i.e., $\partial^\pm c \in V(T_{\partial^- t})$. 
In this case,
\begin{align}
\tilde{F}_{t,c} = \frac{\Gamma_{-c}}{\Gamma_{-t}} e^{A_{P_{\partial^- c \leftarrow \partial^- t}}} \left( 1-e^{A_{C_c}} \right) \, , \label{eqn:Tt_interference}
\end{align}
where
\begin{align}
A_{C_c}=\sum_{e \in C_c} A_e \, , \label{eqn:cycle_affinity}
\end{align}
is the cycle affinity along the cycle $C_c$. 
This term may be interpreted as an Aharonov-Bohm-like interference effect, absent for uniform transition rates.

\begin{figure}[ht]
\begin{center}
\includegraphics[width=0.7 \columnwidth]{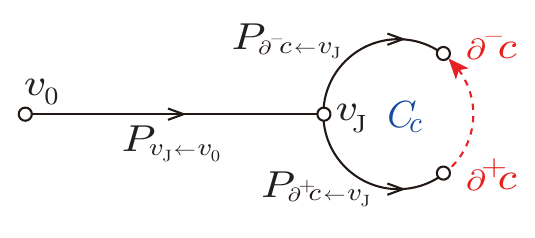}
\caption{
Schematic picture of a twisted cycle. 
Solid lines represent paths connecting specific nodes depicted by empty circles. 
The twisted cycle consists of a cycle $C_c$ and a path connecting the root node $v_0$ to the cycle. 
}
\label{fig:twistedcycle}
\end{center}
\end{figure}
\revis{
In our analysis, the minimization of the distance between the divergence-free current ${\bm j}^\perp$ and the gradient-flow-like current ${\bm j}(\phi)$ in (\ref{eqn:Sigma_w}) naturally leads to both the standard and twisted cycle matrices. 
Although the analogy between the gauge potential and the affinity~\cite{Polettini2012} suggests a connection to a tight-binding Hamiltonian with a gauge potential, a direct counterpart of the twisted cycle is not manifest in the elementary formulation, since the twisted cycle is defined in edge space rather than in the node space where the wave function is defined. 
Figure~\ref{fig:twistedcycle} schematically illustrates a twisted cycle, 
which consists of a cycle $C_c$ and a path connecting the root node to the cycle, 
$P_{v_{\rm J} \leftarrow v_0}$, where $v_{\rm J}$ is a junction node. 
The twisted cycle matrix is, 
\begin{align}
\tilde{B}_{c,t} = 
\frac{\Gamma_{-c}}{\Gamma_{-t}}
e^{A_{P_{\partial^- c \leftarrow \partial^- t}}}
\left \{  
\begin{array}{cc} 
e^{A_{C_c}}-1 & ( t \in P_{v_{\rm J} \leftarrow v_0} ) \\
e^{A_{C_c}} & ( t \in P_{\partial^+ c \leftarrow v_{\rm J}} ) \\
-1 & ( t \in P_{\partial^- c \leftarrow v_{\rm J}} ) \\
\end{array} 
\right .
 \, , \label{eqn:twisted_cycle_eqn}
\end{align} 
and $\tilde{B}_{c,c}=1$, where the cycle affinity $A_{C_c}$ is assumed to be finite. 
Any of the three paths, $P_{v_{\rm J} \leftarrow v_0}$, $P_{\partial^+ c \leftarrow v_{\rm J}}$, and $P_{\partial^- c \leftarrow v_{\rm J}}$, may be empty if their endpoints coincide. 
For example, when $v_0=v_{\rm J}$, $P_{v_{\rm J} \leftarrow v_0}=\emptyset$. 
An empty path contains no twig, and hence the corresponding case in (\ref{eqn:twisted_cycle_eqn}) is absent.
In addition to the phase-accumulation-like factor along the path, $e^{A_{P_{\partial^- c \leftarrow \partial^- t}}}$, which is anticipated from the gauge-potential analogy, a characteristic consequence of finite cycle affinity is the emergence of a path connecting the root node and the cycle whose weight is proportional to $e^{A_{C_c}}-1$. 
The twisted cycle defined in this way is almost orthogonal to the twisted incidence matrix, except at the root node (Appendix \ref{app:Tellegen_tDtB}), 
\begin{align}
\left ( \tilde{\bm D} \tilde{\bm B}^{\sf{T}} \right )_{v,c} = \delta_{v,v_0} \Gamma_{-c} e^{A_{P_{\partial^- c \leftarrow v_0}}} \left ( e^{A_{C_c}} - 1 \right ) \, , 
\label{eqn:Tellegen_tDtB}
\end{align} 
from which one can also confirm (\ref{eqn:Tellegen_t}). 
}

\subsection{Duality transformation: edge to cycle currents}
\label{sec:Duality_transformation}

The constraint (\ref{eqn:set_p1}) is satisfied by setting
\begin{align}
{\bm \phi} = \left( {\bm n}^{\rm st} {\bm 1}^{\sf{T}} - {\bm I}_{|V|} \right) \tilde{\bm S} {\bm \nu}  \, ,
\label{eqn:phi}
\end{align}
where ${\bm \nu} \in \mathbb{R}^{|E(T)|}$. 
Substituting this expression into (\ref{eqn:j_D_twist}), the edge current vector ${\bm j}({\bm \phi}) \in \mathbb{R}^{|E|}$ is expressed in terms of a vector in the twisted cut space, $\tilde{\bm Q}^{\sf{T}} {\bm \nu}$, as
\begin{align}
{\bm j}({\bm \phi}) = {\bm \Phi}^{-1} \tilde{\bm Q}^{\sf{T}} {\bm \nu} \, , 
\qquad
{\bm \Phi} = {\bm I}_{|E|} - \frac{ {\bm j} \left( {\bm n}^{\rm st} \right) {\bm \mu}^{\sf{T}} }{1 + {\bm \mu}^{\sf{T}} {\bm j} \left( {\bm n}^{\rm st} \right) } \, .
\label{eqn:Phi_matrix}
\end{align}
Here, ${\bm \mu} \in \mathbb{R}^{|E|}$ is defined by $\mu_c =0$ for $c \in E(T^*)$ and
\begin{align}
\mu_t = \left( \tilde{\bm S}^{\sf{T}} {\bm 1} \right)_t = \sum_{v \in V \left( T_{\partial^- t} \right) } \frac{e^{A_{P_{v \leftarrow \partial^- t}}}}{\Gamma_{-t}} \, ,
\end{align}
for $t \in E(T)$. 
For uniform transition rates, $\Gamma_{\pm t}=1$, this reduces to the number of nodes contained in the subtree $T_{\partial^- t}$ as, 
$\mu_t = \left| V \left( T_{\partial^- t} \right) \right|$.

In the following, we transform edge currents into cycle currents, in analogy with the vortex--charge duality~\cite{Nagaosa1999}. 
The procedure follows Ref.~\cite{Utsumi2024}, and we briefly outline the derivation below.
Equation (\ref{eqn:Sigma_w}) is equivalently written as
\begin{align}
\frac{\langle \! \langle w \rangle \! \rangle^2}{\langle \! \langle w^2 \rangle \! \rangle}
= \inf_{ {\bm j}^{\perp} \in J^{(1)} } \sup_{ {\bm J} \in \mathbb{R}^{|E|}}
\left( - {\bm J}^{\sf{T}} {\bm G} {\bm J} + 2 {\bm J}^{\sf{T}} {\bm j}^\perp \right) \, ,
\label{eqn:sig_vortex}
\end{align}
subject to the constraint, 
${\bm \Phi}^{-1 \, \sf{T}} {\bm J} \in \ker \tilde{\bm Q}$. 
By exploiting (\ref{eqn:imBTequivKerD}) and (\ref{eqn:imtBTequivKertQ}), 
we parametrize vectors in the kernels, $\ker {\bm D}$ and $\ker \tilde{\bm Q}$,  as
${\bm j}^{\perp} = {\bm B}^{\sf{T}} {\bm f}$
and
${\bm \Phi}^{-1 \, \sf{T}} {\bm J} = \tilde{\bm B}^{\sf{T}} {\bm A}$,
where ${\bm f},{\bm A} \in \mathbb{R}^{|E(T^*)|}$ are cycle (mesh) current vectors.

Then, the right-hand side of (\ref{eqn:sig_vortex}) is rewritten as
\begin{align}
\inf_{ {\bm f} \in F } \sup_{ {\bm A} \in \mathbb{R}^{|E(T^*)|}} 
\left( - {\bm A}^{\sf{T}} {\bm G}_2 {\bm A} 
+ 2 {\bm A}^{\sf{T}} \tilde{\bm B} {\bm \Phi} {\bm B}^{\sf{T}} {\bm f} \right) \, ,
\label{eqn:exp_1chain}
\end{align}
where
\begin{align}
{\bm G}_2  &= \tilde{\bm B} {\bm \Phi} {\bm G} {\bm \Phi}^{\sf{T}} \tilde{\bm B}^{\sf{T}} \in {\mathbb R}^{|E(T^*)| \times |E(T^*)|} \, , \label{eqn:G2} \\
F &= \left \{ {\bm f} \in \mathbb{R}^{|E(T^*)|} \;\middle|\;
{\bm d}^{\sf{T}} {\bm B}^{\sf{T}} {\bm f}  = \langle \! \langle w \rangle \! \rangle = {\bm d}^{\sf{T}} {\bm j} \left( {\bm n}^{\rm st} \right)
\right \} \, .
\label{eqn:sig_2chain}
\end{align}
We find the maximizing ${\bm A}$ from the stationary condition with respect to ${\bm A}$, 
${\bm G}_2 {\bm A} = \tilde{\bm B} {\bm \Phi} {\bm B}^{\sf{T}} {\bm f}$,
which results in
\begin{align}
\frac{\langle \! \langle w \rangle \! \rangle^2}{\langle \! \langle w^2 \rangle \! \rangle} 
=& \inf_{ {\bm f} \in F} {\rm SNR}^2[{\bm f}] 
\, ,  \label{eqn:snr_inf} \\
{\rm SNR}^2[{\bm f}]
=& {\bm f}^{\sf{T}} {\bm B} {\bm \Phi}^{\sf{T}} \tilde{\bm B}^{\sf{T}} {\bm G}_{2}^{-1} \tilde{\bm B} {\bm \Phi} {\bm B}^{\sf{T}} {\bm f} \, .
\label{eqn:snr2}
\end{align}
The original problem reduces to a quadratic optimization problem over the (twisted) cycle space. 
Further calculations of the matrices ${\bm G}_{2}$ and $\tilde{\bm B} {\bm \Phi} {\bm B}^{\sf{T}}$ are presented in Appendix~\ref{app:matrices}. 
A deviation from the optimal point leads to a larger second cumulant, resulting in a TUR~\cite{Utsumi2024}.

\section{applications}
\label{sec:examples}

\subsection{A graph with two cycles}
\label{sec:two_cycles}

\begin{figure}[ht]
\begin{center}
\includegraphics[width=1 \columnwidth]{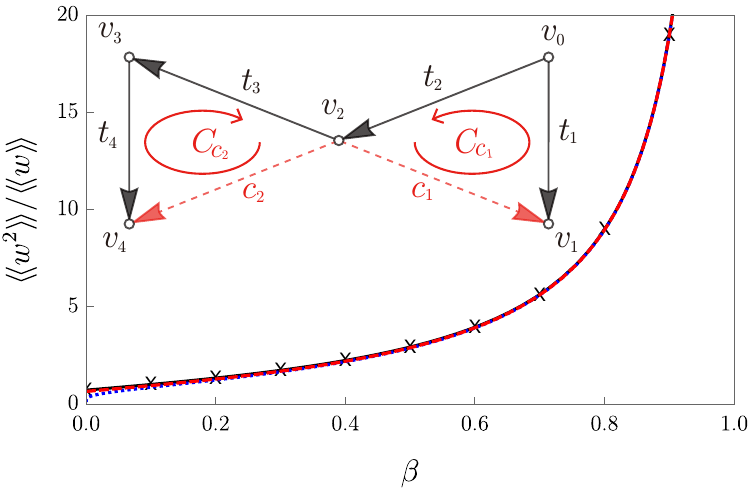}
\caption{
Dependence of the Fano factor on $\beta$. 
The inset shows a graph with two cycles $C_{c_1}$ and $C_{c_2}$. 
The current is measured at the chord $c_2$. 
Crosses indicate the results of second-order perturbation theory~\cite{FlindtPRB2010}. 
The dashed and dotted lines indicate the pseudo-entropy bound $2 \langle \! \langle w \rangle \! \rangle / \sigma_{\rm pseudo}$~\cite{Shiraishi2021} and the mixed bound $2 \langle \! \langle w \rangle \! \rangle / \sigma_{\rm mix}$~\cite{Pal2021}. 
}
\label{fig:fano2cycles_inset}
\end{center}
\end{figure}

As an example, we consider a graph containing two cycles (inset of Fig.~\ref{fig:fano2cycles_inset}). 
Solid and dotted arrows represent twigs and chords, respectively. 
The root node $v_0$ lies on the cycle $C_{c_1}$. 
The two cycles $C_{c_1}$ and $C_{c_2}$ share a node $v_2$ but are edge-disjoint.
We take the transition rates as
\begin{align}
\Gamma_{\pm t_n}=\Gamma_{\pm c_1}=\Gamma_{c_2}=1 \, , 
\;\;\;\;
\Gamma_{-c_2} = \beta \, ,
\end{align}
where $n=1,2,\dots,4$.
Then the cycle affinities are $A_{C_{c_1}}=0$ and $A_{C_{c_2}}=-\ln \beta$. 

Since the second cycle $C_{c_2}$ is located downstream of $t_2$, the cycle affinity $A_{C_{c_2}}$ induces an interference-like term:
\begin{align}
\tilde{F}_{t_2,c_2} =& \frac{\Gamma_{-c_2}}{\Gamma_{-t_2}} e^{A_{P_{\partial^- c_2 \leftarrow \partial^- t_2}}} (1-e^{A_{C_{c_2}}}) \nonumber \\
=&  \beta e^{A_{P_{v_4 \leftarrow v_2}}} (1-1/\beta)=\beta-1 \, . 
\end{align}
Other components of the twisted cycle matrix can be calculated in the same way: 
\begin{align}
\tilde{\bm B} =
\begin{bmatrix}
-1 &   1 &   0 &   0 &   1 &   0 \\
 0 & 1-\beta &  -\beta &  -\beta &   0 &   1  
\end{bmatrix}
\, ,
\end{align}
where the rows correspond to the cycles $C_{c_1}$ and $C_{c_2}$, and the columns to the arcs $t_1$, $t_2$, $t_3$, $t_4$, $c_1$, and $c_2$.

The cycle affinity modifies the Gram matrix as
\begin{align}
\tilde{\bm B} {\bm B}^{\sf{T}} =
\begin{bmatrix}
       3 &   0  \\
 1-\beta & 1+2 \beta \\
\end{bmatrix} \in {\mathbb R}^{|E(T^*)| \times |E(T^*)|} \, , \label{eqn:gram_2cycles}
\end{align}
where the rows and columns correspond to the cycles $C_{c_1}$ and $C_{c_2}$. 
In the absence of affinity ($\beta=1$), the $(c,c')$ element represents the number of arcs shared by the cycles $C_c$ and $C_{c'}$ with the same orientation minus those with the opposite orientation. 
Therefore, the diagonal components correspond to the lengths of the cycles, which are $3$, while the off-diagonal components vanish since the two cycles do not share any arcs. 
Equation (\ref{eqn:gram_2cycles}) implies that the cycle affinity induces an effective overlap between distant cycles and modifies their effective lengths. 

Figure~\ref{fig:fano2cycles_inset} shows the $\beta$ dependence of the Fano factor of the current measured at $c_2$, i.e., $d_e = \delta_{e,c_2}$ (Appendix~\ref{eqn:2cycleG}):
\begin{align}
\frac{ \langle \! \langle w^2 \rangle \! \rangle }{ \langle \! \langle w \rangle \! \rangle } = \frac{46 + 214 \beta + 139 \beta^2 + 51 \beta^3}{(8+7\beta)^2 (1-\beta)}
\, . \label{eqn:fano_2cycleG}
\end{align}
In the limit of large forward bias $\beta \to 0$, the Fano factor is sub-Poissonian, $23/32$, while in the equilibrium limit $\beta \to 1$, it diverges. 
For comparison, we show the results calculated using second-order perturbation theory (crosses)~\cite{FlindtPRB2010}, which agree well with (\ref{eqn:fano_2cycleG}).

The lower bound of the Fano factor can be estimated using TUR, 
as $\langle \! \langle w^2 \rangle \! \rangle/\langle \! \langle w \rangle \! \rangle \geq 2 \langle \! \langle w \rangle \! \rangle /\sigma$ for $\langle \! \langle w \rangle \! \rangle>0$~\cite{Horowitz2020}. 
The dashed and dotted lines indicate the bounds associated with the pseudo-entropy
$2 \langle \! \langle w \rangle \! \rangle / \sigma_{\rm pseudo}$~\cite{Shiraishi2021}
and with the mixed description of entropy and activity
$2 \langle \! \langle w \rangle \! \rangle / \sigma_{\rm mix}$~\cite{Pal2021},
with $\sigma_{{\rm pseudo}({\rm mix})} = {\bm d}^{{\rm pseudo}({\rm mix}){\sf T}} {\bm j} \left ( {\bm n}^{\rm st} \right )$: 
\begin{align}
d_e^{\rm mix} =& \left \{
\begin{array}{cc}
\ln \frac{\Gamma_e n_{\partial^+ e}}{\Gamma_{-e} n_{\partial^- e}} & (e \in E_{\rm b}) \\
2 \Gamma_e n_{\partial^+ e} & (e \in E_{\rm uni})
\end{array}
\right . \, , \label{eqn:def_mix} \\
d_e^{\rm pseudo} =& 2 j^{\rm st}_e/g^{\rm st}_e \, . 
\end{align}
Since all transition processes are bidirectional for the present case, 
$\sigma_{\rm mix}= - \langle \! \langle w \rangle \! \rangle \ln \beta$ coincides with the entropy production rate. 
For the explicit form of the pseudo-entropy, see (\ref{eqn:pseudo_ent_2cycles}). 
The two bounds are rather tight, except for small $\beta$, where the affinity $-\ln \beta$ diverges and the mixed bound approaches 0.

\subsection{Exponentially branched logically-irreversible Brownian computation}
\label{sec:LIBC}

\begin{figure}[ht]
\begin{center}
\includegraphics[width=1 \columnwidth]{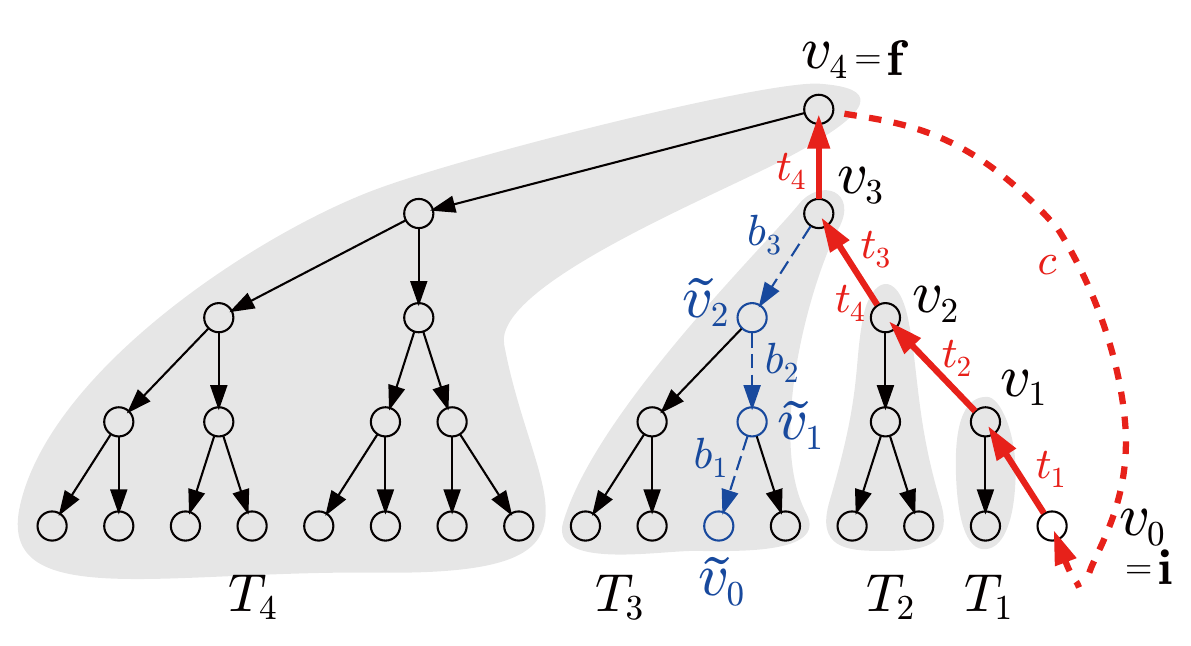}
\caption{
State-transition diagram illustrating a Brownian computation model with a reset process. 
Solid and dashed arcs constitute a directed rooted spanning tree rooted at $v_0={\rm i}$, which is the initial state. 
Thick solid arcs indicate the intended computation path from the initial state $v_0={\rm i}$ to the output state $v_\ell={\rm f}$. 
The path length $\ell$ ($\ell=4$) corresponds to the minimum number of computational steps. 
The dotted arc is a chord representing the reset process. 
The $\alpha$-fold backward branchings ($\alpha=2$) represent logical irreversibility, and shaded subtrees rooted at the logical states $v_d$ ($d=1, \dots, \ell$) include their extraneous predecessors as well as the logical states themselves.
Dashed arcs indicate excursions among extraneous predecessors leading to another possible input state $\tilde{v}_0$. 
}
\label{fig:gTree_comp}
\end{center}
\end{figure}

The stochastic dynamics of the Brownian computation process is modeled as a Brownian search process on a tree-like graph~\cite{Bennett1982,Norton2013}. 
Figure~\ref{fig:gTree_comp} shows an example: 
a graph with the structure of a complete $\alpha$-ary tree of height $\ell$~\cite{Cormen_Thomas_H_2022-04-05}, 
illustrated here for $\alpha=2$ and $\ell=4$. 
The nodes represent logical states: the nodes at the lowest level correspond to possible input states, and the node at the highest level corresponds to a unique output state, denoted by $v_\ell={\rm f}$. 
The computation proceeds from the lowest level toward the highest level, and the $\alpha$-fold backward branching represents logical irreversibility.

Solid and dashed arcs form a directed rooted spanning tree $T=(V,E(T))$. 
The computation proceeds from the initial state $v_0={\rm i}$, taken as the root node from the bottom nodes, to the final state $v_\ell={\rm f}$. 
An intended computation path is given by $v_0 \to v_1 \to \cdots \to v_\ell$ [thick solid arcs], with the corresponding transition sequence written in right-to-left order as, 
\begin{align}
P_{v_\ell \leftarrow v_0} = \left( t_{\ell}, t_{\ell-1}, \dots, t_1 \right).
\end{align}
where $t_d = (v_d \leftarrow v_{d-1})$ denotes the transition from $v_{d-1}$ to $v_d$ ($d=1,2,\dots,\ell$). 
For Brownian Boolean circuits, the path length $\ell$ corresponds to the number of gates~\cite{Utsumi2022computation,Okajima2026}.

At each node on the intended computation path, a subtree $T_d$ rooted at the node $v_d$ is attached [shaded parts in Fig.~\ref{fig:gTree_comp}]. 
The set of nodes is then given by $V = \bigcup_{d=1}^\ell V(T_d) \cup \{ v_0 \}$. 
The nodes, except for the roots of the subtrees $\tilde{v} \in V(T_d) \backslash \{ v_d \}$, represent extraneous predecessors, 
i.e., states that do not belong to the intended computation path. 
We assume that the transition rate in the forward direction is 1, while that in the backward direction is $\beta$:
\begin{align}
\Gamma_{t_d} = \Gamma_{-b} = 1, \qquad \Gamma_{-t_d} = \Gamma_{b} = \beta,
\end{align}
where $t_d \in P_{v_\ell \leftarrow v_0}$ is a twig on the intended computation path, while $b \in \bigcup_{d=1}^\ell E(T_{d})$ is a twig on the subtrees. 
The bias is in the forward direction for $\beta<1$ and in the backward direction for $\beta>1$. 
$\beta=1$ corresponds to the unbiased case discussed in Ref.~\cite{Utsumi2024}.

The dotted arc in Fig.~\ref{fig:gTree_comp} is a chord $c=(v_0 \leftarrow v_{\ell})$ representing the reset process, i.e., $\Gamma_c = 1$ and $\Gamma_{-c} \to 0$. 
The intended computation path $P_{v_\ell \leftarrow v_0}$ and the reset arc $c$ together form a cycle by concatenation, 
\begin{align}
C_c = \left( c, t_{\ell}, t_{\ell-1}, \dots, t_1 \right) ,
\end{align}
where the sequence is read from right to left. 
The reset current is measured at the chord $c$, i.e., $d_e = \delta_{e,c}$.

\begin{figure}[ht]
\begin{center}
\includegraphics[width=0.9 \columnwidth]{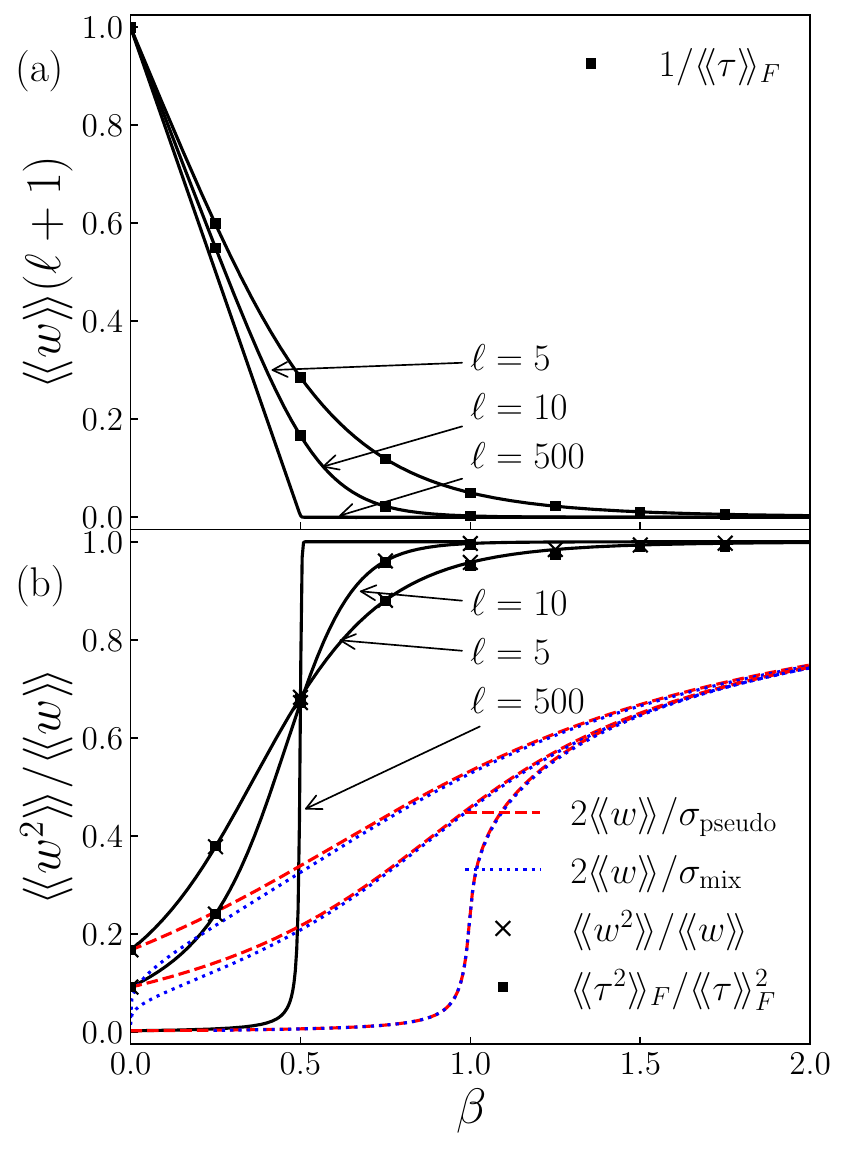}
\caption{
The average reset current (a) and its Fano factor (b) are shown as functions of the backward transition rate $\beta$. 
The solid lines in panels (a) and (b) represent the analytic results (\ref{eqn:aw}) and (\ref{eqn:fano_ana}), respectively, for $\ell=5,10,500$ with branching factor $\alpha=2$.
Filled squares indicate numerical results obtained by the Gillespie algorithm with $10^5$ samples. 
The crosses in panel (b) indicate the result of second-order perturbation theory~\cite{FlindtPRB2010}. 
The dashed and dotted lines in panel (b) indicate the pseudo-entropy bound~\cite{Shiraishi2021} and the mixed bound~\cite{Pal2021}, respectively.
}
\label{fig:cumlants_vs_beta_1resets}
\end{center}
\end{figure}

For the state-transition diagram in Fig.~\ref{fig:gTree_comp}, there is only one cycle, and hence there are no free parameters in (\ref{eqn:snr_inf}). 
The average reset current and the Fano factor are given by (see Appendix \ref{app:LIBC} and Ref.~\cite{SM} for detailed calculations), 
\begin{align}
\langle \! \langle w \rangle \! \rangle = \frac{(1-x)^2}{\Delta} \, ,
\;\;\; \Delta = \tilde{\ell}(1-x)+x \left( x^{\tilde{\ell}}-1 \right) \, , \label{eqn:aw}
\end{align}
\begin{align}
\frac{ \langle \! \langle w^2 \rangle \! \rangle }{\langle \! \langle w \rangle \! \rangle } =& \left[ 
\tilde{\ell}(1-x) \left( 1+x+4 x^{\tilde{\ell}+1} \right) + x \left( x^{\tilde{\ell}}-1 \right) 
\right . \nonumber \\  & \times \left.  
\left( 4+x+x^{\tilde{\ell}+1} \right) \right] \Delta^{-2} 
\, , \label{eqn:fano_ana}
\end{align}
where $\tilde{\ell}=\ell+1$ and $x=\alpha \beta$. 
The result coincides with Ref.~\cite{Utsumi2024} in the absence of affinity ($\beta=1$), 
indicating that the backward bias $\beta$ and the branching factor $\alpha$ play equivalent roles. 
This dependence only on the product $\alpha \beta$ supports the embedding procedure discussed in Ref.~\cite{Okajima2026}. 
Figures \ref{fig:cumlants_vs_beta_1resets}(a) and (b) show the average reset current and the Fano factor as functions of $\beta$, respectively. 
The three solid lines in each panel show the analytic results for $\ell=5, 10$, and $500$ (the branching factor is $\alpha=2$). 

In each panel, squares indicate the numerical results for the reciprocal of the average first-passage time $1/\langle \! \langle \tau \rangle \! \rangle_F$ and the inverse square of the signal-to-noise ratio, 
$\langle \! \langle \tau^2 \rangle \! \rangle_F/\langle \! \langle \tau \rangle \! \rangle_F^2$. 
These approach the average reset current $1/\langle \! \langle \tau \rangle \! \rangle_F \approx \langle \! \langle w \rangle \! \rangle$ and the Fano factor $\langle \! \langle \tau^2 \rangle \! \rangle_F/\langle \! \langle \tau \rangle \! \rangle_F^2 \approx \langle \! \langle w^2 \rangle \! \rangle/\langle \! \langle w \rangle \! \rangle$, respectively, for many resets~\cite{Utsumi2022computation}. 
The crosses in panel (b) indicate the results of second-order perturbation theory~\cite{FlindtPRB2010}. 
All three approaches are consistent with each other. 
This agreement confirms the analytic expressions (\ref{eqn:aw}) and (\ref{eqn:fano_ana}).

In the limit of an infinitely long intended computation path, the average reset current and the Fano factor become
\begin{align}
\lim_{\tilde{\ell} \to \infty} \langle \! \langle w \rangle \! \rangle \tilde{\ell} &= (1-x)\,\theta(1-x) \, , \\
\lim_{\tilde{\ell} \to \infty} \frac{\langle \! \langle w^2 \rangle \! \rangle}{\langle \! \langle w \rangle \! \rangle}  &= \theta(x-1) \, . 
\end{align}
The abrupt change at $x=1$ would correspond to the easy-hard transition in the Brownian computational time complexity~\cite{Okajima2026}. 
When the forward bias is sufficiently strong, i.e., $x<1$, 
the average reset current scales as $\tilde{\ell}^{-1}$, 
$\langle \! \langle w \rangle \! \rangle \approx (1-x)/\tilde{\ell}$
as in a series of Ohmic resistors. 
In contrast, when the forward bias is not strong, $x>1$, 
the reset current decays exponentially with $\tilde{\ell}$, 
$\langle \! \langle w \rangle \! \rangle \approx x^{-{\ell}} (1-1/x)^2$. 
For $x \to 1$, the reset current exhibits diffusive scaling, 
$\langle \! \langle w \rangle \! \rangle \to 2/(\tilde{\ell} \ell)$~\cite{Utsumi2023}. 

The dashed and dotted lines in panel (b) show the pseudo-entropy bound (\ref{eqn:pseudo}) and the mixed bound (\ref{eqn:mix}). 
For large backward transition rates $\beta \gg 1$, they approach unity. 
In the limit of vanishing backward transition rate, $\beta \to 0$, the lower bound of Fano factor estimated from the pseudo-entropy reproduces the exact value 
$\lim_{\beta \to 0} 2 \langle \! \langle w \rangle \! \rangle/ \sigma_{\rm pseudo} = 1/ \tilde{\ell}$. 
In the limit $\ell \to \infty$, the bound approaches zero for $\beta<1$, i.e., $2 \langle \! \langle w \rangle \! \rangle/\sigma_{\rm pseudo} \approx 0$.
As $\beta \to 1^+$, the bound vanishes as $2 \langle \! \langle w \rangle \! \rangle/\sigma_{\rm pseudo} \approx -2/\ln(\beta-1)$, as is also observed for the mixed bound (see Appendix~\ref{app:ssTURs}). 
Although this non-analytic behavior is reminiscent of the easy--hard transition, it occurs at $\beta=1$ and does not coincide with the transition point.

In the mixed bound, $\sigma_{\rm mix}$ consists of entropic and kinetic terms, \revis{see (\ref{eqn:def_mix})}:
$\sigma_{\rm mix} = \sigma_{\rm tot}^{\rm bi} + 2 \langle \! \langle w \rangle \! \rangle$, 
where $\sigma_{\rm tot}^{\rm bi} = \sigma_{\rm env}^{\rm bi}+\sigma_{\rm sys}^{\rm bi}$.  
Here, the environment and system entropy production rates due to bidirectional processes are given by 
\begin{align}
\sigma_{\rm env}^{\rm bi} = \langle \! \langle w \rangle \! \rangle \ell \ln \frac{1}{\beta} \, , 
\quad 
\sigma_{\rm sys}^{\rm bi} = \langle \! \langle w \rangle \! \rangle \ln \frac{\beta^{\tilde{\ell}}-1}{\beta-1} 
\, . \label{eqn:enbi}
\end{align}
They depend on $\alpha$ only through the average reset current $\langle \! \langle w \rangle \! \rangle$. 
In the context of the stochastic thermodynamics of resetting, $\sigma_{\rm tot}^{\rm bi}$ represents the minimal thermodynamic cost required for resetting~\cite{Fuchs2016,Utsumi2022computation,Utsumi2023}. 
In the steady state, $-\sigma_{\rm sys}^{\rm bi}$ coincides with the reset entropy production rate.
In the zero-affinity limit $\beta \to 1$, where the computation time $\langle \! \langle \tau \rangle \! \rangle_F$ grows quadratically ($\alpha=1$) or exponentially ($\alpha>1$) with $\ell$, the heat emission vanishes, $\sigma_{\rm env}^{\rm bi} = 0$. 
For a single run, the negative reset entropy production becomes 
$\sigma_{\rm sys}^{\rm bi} \langle \! \langle \tau \rangle \! \rangle_F \approx \sigma_{\rm sys}^{\rm bi}/\langle \! \langle w \rangle \! \rangle = \ln \tilde{\ell} $. 
It increases logarithmically with the length of the intended computation path~\cite{Utsumi2024}. 
The reset entropy can be interpreted as the entropy associated with the free expansion of a Brownian particle. 
In this sense, the result does not contradict Refs.~\cite{Norton2013,Strasberg2015}.
At the easy-hard transition point, $1/\beta=\alpha$, the environment entropy production reads, 
$ \sigma_{\rm env}^{\rm bi} \langle \! \langle \tau \rangle \! \rangle_F  \approx \ell \ln \alpha $.
This result does not contradict Ref.~\cite{Bennett1982}, which argues that $\alpha$ immediate predecessors would retard the computation unless a thermodynamic force of order $\ln \alpha$ per step is present. 

\revis{
The steady-state entropy production admits a cycle decomposition~\cite{Schnakenberg1976,Andrieux2007}. 
In the present case, where only a single reset cycle is present, the cycle affinity $A_{C_c}$ diverges because of the unidirectional reset transition. 
Nevertheless, the entropy production associated with the bidirectional transitions, see (\ref{eqn:def_mix}), can be written in a similar form by using the fact that the steady-state edge current is the same on every edge of the cycle, 
$j_{e}(n_v^{\rm st}) = \langle \! \langle w \rangle \! \rangle$ for $e \in C_c$: 
\begin{align}
\sigma_{{\rm bi}}^{\rm tot} 
=&
\sum_{e \in C_c \backslash \{c\}} j_e(n_v^{\rm st}) 
\ln \frac{\Gamma_e n_{\partial^+ e}^{\rm st}}{\Gamma_{-e} n_{\partial^- e}^{\rm st}} 
\nonumber \\
=&
\langle \! \langle w \rangle \! \rangle 
\sum_{e \in C_c \backslash \{c\}} A_{e} 
+ 
\langle \! \langle w \rangle \! \rangle 
\sum_{e \in C_c \backslash \{c\}} \ln \frac{n_{\partial^+ e}^{\rm st}}{n_{\partial^- e}^{\rm st}} 
\, .
\end{align}
Here, the partial cycle affinity is 
$
\sum_{e \in C_c \backslash \{c\}} A_{e} 
=
\sum_{d=1}^\ell A_{t_d} 
=
- \ell \ln \beta 
$. 
The system-entropy contribution can be evaluated as a telescoping sum along the path $C_c \backslash \{c\}$: 
$
\sum_{e \in C_c \backslash \{c\}} \ln ( n_{\partial^+ e}^{\rm st}/n_{\partial^- e}^{\rm st}) 
= 
\ln (n_{v_0}^{\rm st}/n_{v_\ell}^{\rm st})
= \ln [(\beta^{\tilde{\ell}}-1)/(\beta-1)]$. 
Although the reset current $\langle \! \langle w \rangle \! \rangle$ changes its asymptotic behavior at the easy--hard transition point, $\beta=1/\alpha$, 
the ratio 
$\sigma_{{\rm bi}}^{\rm tot} /\langle \! \langle w \rangle \! \rangle$ 
entering the TUR bound for the Fano factor is independent of $\alpha$. 
This explains why the TUR bound does not capture the easy--hard transition point. 
}

In the above discussion, we considered a schematic tree--graph representation of the Brownian computation process. 
In practical implementations of token-based Brownian circuits~\cite{Lee2016}, the finite affinity arises from the difference in the forward and backward transition rates of CJoin~\cite{Utsumi2022computation}. 
Local cycles may arise from concurrency, i.e., from the commutativity of independent firings of CJoins. 
Such cycles ideally carry no affinity, as the transition rates are identical and independent of the firing order. 
Therefore, they are expected not to interfere with the reset cycle. 
The token-based Brownian circuit can be made logically reversible by adopting a CJoin with memory~\cite{Utsumi2026}, which allows distinguishing different computation paths but requires resetting the memory registers after the computation.

\section{Conclusion}
\label{sec:conclusion}

We introduce circuit matrices twisted by a finite affinity that preserve the orthogonality between the twisted cycle and cut spaces. 
The twisted cycle matrix exhibits an interference-like effect associated with finite cycle affinity. 
This framework allows us to perform a duality transformation from the twisted cut space to the twisted cycle space, 
which is particularly useful for state-transition diagrams with a small number of cycles. 
We apply the formalism to a graph with two edge-disjoint cycles and show that a finite cycle affinity induces an interference-like effect that appears in the Gram matrix as an effective overlap and a change in cycle length. 
For a model of logically irreversible Brownian computation on an $\alpha$-ary tree-like graph, we derive an analytic expression for the Fano factor of the reset current, which exhibits a transition from zero to unity when the thermodynamic force equals $\ln \alpha$ in the limit of an infinitely long computation path.
This transition corresponds to an easy–hard transition from linear to exponential scaling of the computation time with the length of the computation path~\cite{Okajima2026}.
Our result provides an analytic perspective on this transition and connects it to the thermodynamic cost of logically irreversible computation~\cite{Bennett1982}.

\begin{acknowledgments}
We thank Sho Nakade for valuable comments. 
This work was supported by JSPS KAKENHI Grants No. 24K00547 and No. 24H01336. 
\end{acknowledgments}

\appendix

\section{Proof of Eq.~(\ref{eqn:Q_twist})}
\label{app:modified_cutset_matrix}

By using (\ref{eqn:D_twist}) and (\ref{eqn:gpathmat}), (\ref{eqn:Q_twist}) is calculated as, 
\begin{align}
\tilde{Q}_{t,e} =& \frac{\Gamma_{-e}}{\Gamma_{-t}} \biggl ( 
- e^{A_e + A_{P_{\partial^+ e \leftarrow \partial^- t}} } \mathbbm{1}_{V(T_{\partial^- t})} (\partial^+ e)
\nonumber \\ 
&+ e^{A_{P_{\partial^- e \leftarrow \partial^- t}}} \mathbbm{1}_{V(T_{\partial^- t})} (\partial^- e) 
\biggr ) \, , 
\label{eqn:Q_twist_1}
\end{align} 
where we have assumed that $\Gamma_{\pm t} >0$ for all $t \in E(T)$, which ensures that the expression is well defined.

For a twig $e = t' \in E(T)$, this reduces to
\begin{align}
\tilde{Q}_{t,t'} =& \frac{\Gamma_{-t'}}{\Gamma_{-t}} \left ( -\mathbbm{1}_{V(T_{\partial^- t})} (\partial^+ t') + \mathbbm{1}_{V(T_{\partial^- t})} (\partial^- t') \right )
\nonumber \\ 
& \times e^{A_{P_{\partial^- t' \leftarrow \partial^- t}}} = \delta_{t,t'} \, ,
\end{align}
where we used the additivity of the affinity along concatenated paths,
$A_{t'} + A_{P_{\partial^+ t' \leftarrow \partial^- t}} = A_{P_{\partial^- t' \leftarrow \partial^- t}}$. 

For a chord \(e=c \in E(T^*)\),
\begin{align}
\tilde{Q}_{t,c} =& \tilde{F}_{t,c} 
= 
\frac{\Gamma_{-c}}{\Gamma_{-t}} 
\biggl ( e^{A_{P_{\partial^- c \leftarrow \partial^- t}}} \mathbbm{1}_{V(T_{\partial^- t})} (\partial^- c)
\nonumber \\ 
& - e^{A_c + A_{P_{\partial^+ c \leftarrow \partial^- t}} } \mathbbm{1}_{V(T_{\partial^- t})} (\partial^+ c) \biggr ) 
\, .
\label{eqn:interference_matrix}
\end{align} 
%
For $\partial^+ c \in V(T_{\partial^- t})$ and $\partial^- c \notin V(T_{\partial^- t})$, 
which means that $t \in C_c$ [Fig.~\ref{fig:graph_interference}(a)], 
only the second term on the right-hand side of (\ref{eqn:interference_matrix}) remains, 
and we obtain (\ref{eqn:Tt_para}). 
For $\partial^- c \in V(T_{\partial^- t})$ and $\partial^+ c \notin V(T_{\partial^- t})$, 
which means that $-t \in C_c$ [Fig.~\ref{fig:graph_interference}(b)], 
only the first term on the right-hand side of (\ref{eqn:interference_matrix}) remains, and we obtain (\ref{eqn:Tt_antipara}). 
For $\partial^\pm c \in V(T_{\partial^- t})$, 
which means that $C_c$ is located downstream of $t$ [Fig.~\ref{fig:graph_interference}(c)], both terms remain. 
We then use the additivity of the affinity along concatenated paths,
\begin{align}
A_{C_c} = A_c + A_{P_{ \partial^+ c \leftarrow \partial^- t} } - A_{P_{\partial^- c \leftarrow \partial^- t}} \, , 
\end{align}
and obtain (\ref{eqn:Tt_interference}).

\section{Proof of Eq.~(\ref{eqn:Tellegen_tDtB})}
\label{app:Tellegen_tDtB}

\revis{
The left-hand side of (\ref{eqn:Tellegen_tDtB}) can be written as
\begin{align}
\left ( \tilde{\bm D} \tilde{\bm B}^{\sf{T}} \right )_{v,c} = \sum_{e \in \delta^+ v} \Gamma_{e} \tilde{B}_{c,e} - \sum_{e \in \delta^- v} \Gamma_{-e} \tilde{B}_{c,e} \, ,
\label{eqn:app_Tellegen_tDtB_1}
\end{align} 
where $\delta^\pm v = \left \{ e \middle | \partial^\pm e =v \wedge e \in E \right \}$. 
The remaining calculation can be carried out separately for each case. 
At the junction node $v=v_J$, 
when one incoming twig and two outgoing twigs are present, 
$\delta^+ v = \{ t_{\rm oF} , t_{\rm oB} \}$ and 
$\delta^- v = \{ t_{\rm i} \}$, 
where 
$t_{\rm oF} \in P_{\partial^+ c \leftarrow v_{\rm J}}$ 
and
$t_{\rm oB} \in P_{\partial^- c \leftarrow v_{\rm J}}$, 
(\ref{eqn:app_Tellegen_tDtB_1}) is calculated as, 
\begin{align}
&\Gamma_{t_{\rm oF}} \tilde{B}_{c,t_{\rm oF}} + \Gamma_{t_{\rm oB}} \tilde{B}_{c,t_{\rm oB}} - \Gamma_{-t_{\rm i}} \tilde{B}_{c,t_{\rm i}} \nonumber \\
=&
\Gamma_{-c} e^{A_{P_{\partial^- c \leftarrow \partial^- t_{\rm oF}}}+A_{ t_{\rm oF} }} e^{A_{C_c}} - \Gamma_{-c} e^{A_{P_{\partial^- c \leftarrow \partial^- t_{\rm oB}}}+A_{ t_{\rm oB} }}
\nonumber \\
&- \Gamma_{-c} e^{A_{P_{\partial^- c \leftarrow v}}} ( e^{A_{C_c}} -1) =0
\, . 
\end{align} 
%
At the root node, $v=v_0$, when an outgoing twig is present $\delta^+ v = \{ t_{\rm o} \}$ and no incoming chord is present $\delta^- v = \emptyset$, 
(\ref{eqn:app_Tellegen_tDtB_1}) reads, 
\begin{align}
& \Gamma_{t_{\rm o}} \tilde{B}_{c,t_{\rm o}} =
\Gamma_{-c}
e^{A_{t_{\rm o}}+A_{P_{\partial^- c \leftarrow \partial^- t_{\rm o}}}}
\left ( e^{A_{C_c}} - 1 \right )
\nonumber \\
=&
\Gamma_{-c} e^{A_{P_{\partial^- c \leftarrow v}}} \left ( e^{A_{C_c}} - 1 \right )
\, . 
\end{align} 
%
For a node on the path connecting the root node to the cycle,
$v \in P_{v_{\rm J} \leftarrow v_0}$ with $v \neq v_{\rm J}, v_0$, 
when one outgoing twig and one incoming twig are present,
$\delta^+ v = \{ t_{\rm o} \}$ and $\delta^- v = \{ t_{\rm i} \}$, 
(\ref{eqn:app_Tellegen_tDtB_1}) reads, 
\begin{align}
\Gamma_{t_{\rm o}} \tilde{B}_{c,t_{\rm o}} - \Gamma_{-t_{\rm i}} \tilde{B}_{c,t_{\rm i}} =0
\, . 
\end{align} 
The remaining cases are evaluated analogously, completing the proof of (\ref{eqn:Tellegen_tDtB}).
}

\section{Explicit expressions of matrices in (\ref{eqn:snr2})}
\label{app:matrices}

By utilizing the cycle decomposition of the steady-state current~\cite{Schnakenberg1976},
\begin{align}
{\bm j} \left ( {\bm n}^{\rm st} \right ) = {\bm B}^{\sf{T}} {\bm i}^{\rm st} \, . 
\end{align}
Then by using (\ref{eqn:Phi_matrix}), we obtain 
\begin{align}
\tilde{\bm B} {\bm \Phi} {\bm B}^{\sf{T}} =& \tilde{\bm B} {\bm B}^{\sf{T}} \left [ {\bm I}_{|E(T^*)|} -  \frac{ {\bm i}^{\rm st} \left ( {\bm B} {\bm \mu} \right )^{\sf{T}} } {1 + {{\bm i}^{\rm st}}^{\sf{T}} {\bm B} {\bm \mu}} \right ] \, , 
\label{eqn:Gram_Phi}
\end{align}
where the Gram matrix $ \tilde{\bm B} {\bm B}^{\sf{T}} \in \mathbb{R}^{|E(T^*)| \times |E(T^*)|} $
and the vector in the cycle space ${\bm B} {\bm \mu} \in \mathbb{R}^{|E(T^*)|}$ are expressed as 
\begin{align}
\left ( \tilde{\bm B} {\bm B}^{\sf{T}} \right )_{c,c'} =& \sum_{t \in E(T) \cap \left( C_{c'} \cup \overline{C_{c'}} \right) } (-1)^{\mathbbm{1}_{C_{c'}}(t)}\tilde{F}_{t,c} + \delta_{c,c'} \, , 
\label{eqn:Gram_twist}
\\
\left ( {\bm B} {\bm \mu} \right )_c =& \sum_{ t \in E(T) \cap \left( C_c \cup \overline{C_c} \right) } (-1)^{\mathbbm{1}_{C_{c}}(-t)} {\mu}_{t} \, .  
\label{eqn:Bmu}
\end{align}
Here $t \in E(T) \cap \left( C_c \cup \overline{C_c} \right)$ indicates that the twig $t$ lies on the cycle $C_c$, irrespective of its orientation. 

By substituting (\ref{eqn:Phi_matrix}) into (\ref{eqn:G2}), we obtain 
\begin{align}
{\bm G}_2 =& \tilde{\bm B} {\bm G} \tilde{\bm B}^{\sf{T}} 
+ \frac{ {\bm \mu}^{\sf{T}} {\bm G} {\bm \mu} }{\left ( 1 + {{\bm i}^{\rm st}}^{\sf{T}} {\bm B} {\bm \mu} \right )^2} 
\tilde{\bm B} {\bm B}^{\sf{T}} {\bm i}^{\rm st} \left ( \tilde{\bm B} {\bm B}^{\sf{T}} {\bm i}^{\rm st} \right )^{\sf{T}}
\nonumber \\
&-
\frac{
   \tilde{\bm B} {\bm B}^{\sf{T}} {\bm i}^{\rm st} \left ( \tilde{\bm B} {\bm G} {\bm \mu} \right )^{\sf{T}} 
   + \tilde{\bm B} {\bm G} {\bm \mu} \left ( \tilde{\bm B} {\bm B}^{\sf{T}} {\bm i}^{\rm st} \right )^{\sf{T}}
}{1 + {{\bm i}^{\rm st}}^{\sf{T}} {\bm B} {\bm \mu}}
 \, . 
\label{eqn:G2explicit}
\end{align}
The weighted cycle Laplacian $\tilde{\bm B} {\bm G} \tilde{\bm B}^{\sf{T}} \in \mathbb{R}^{|E(T^*)| \times |E(T^*)|}$, 
the cycle-space vector $\tilde{\bm B} {\bm G} {\bm \mu} \in \mathbb{R}^{|E(T^*)|}$, 
and the weighted squared norm of ${\bm \mu}$, ${\bm \mu}^{\sf{T}} {\bm G} {\bm \mu} \in \mathbb{R}$ 
are expressed as
\begin{align}
\left ( \tilde{\bm B} {\bm G} \tilde{\bm B}^{\sf{T}} \right )_{c,c'} =& \sum_{t \in E(T)} \tilde{F}_{t,c} \tilde{F}_{t,c'} g_t + g_c \delta_{c,c'} \, , 
\label{eqn:cycleLaplacian_bare}
\\
\left ( \tilde{\bm B} {\bm G} {\bm \mu} \right )_{c} =& - \sum_{t \in E(T)} \tilde{F}_{t,c} g_{t} \mu_t \, ,  
\label{eqn:BGmu}
\\
{\bm \mu}^{\sf{T}} {\bm G} {\bm \mu} =& \sum_{t \in E(T)} g_t \mu_t^2 \, , 
\label{eqn:mu_sq_norm}
\end{align}
where $c,c' \in E(T^*)$.
The matrix elements of ${\bm G}_2$ and $\tilde{\bm B} {\bm \Phi} {\bm B}^{\sf{T}}$ are expressed in terms of $g_e$, $\tilde{F}_{t,c}$, and $\mu_t$, which makes the dependence on edge weights and the cycle structure explicit and facilitates further analytical evaluation.

\section{Detailed calculations of Sec.~\ref{sec:two_cycles}}
\label{eqn:2cycleG}

The steady-state probability can be calculated using the Kirchhoff-Hill theorem~\cite{Schnakenberg1976,Weidlich1978}. 
The node state probability is $n^{\rm st}_{v} = W_{v}/Z$, where $W_v$ denotes the total weight of directed spanning trees directed toward $v$, 
\begin{align}
W_{v_n} =&
\left\{
\begin{array}{ll}
W_{v_n}^{(c_1)} W_{v_2}^{(c_2)} , & (n=0,1,2) \\
W_{v_2}^{(c_1)} W_{v_n}^{(c_2)} , & (n=3,4)
\end{array}
\right. , 
\end{align}
\begin{align}
W_{v_0}^{(c_1)} =& W_{v_1}^{(c_1)} = W_{v_2}^{(c_1)} = W_{v_4}^{(c_2)} = 3 \, ,
\\
W_{v_2}^{(c_2)} =& 2 \beta + 1 \, , \quad W_{v_3}^{(c_2)} = \beta + 2 \, .
\end{align}
The normalization factor is then given by $Z= \sum_{v \in V} W_{v} = 3 (7\beta+8)$.

The incidence matrix is, 
\begin{align}
{\bm D} =
\begin{bmatrix}
   1 &   1 &   0 &   0 &   0 &   0   \\
  -1 &   0 &   0 &   0 &  -1 &   0   \\
   0 &  -1 &   1 &   0 &   1 &   1   \\
   0 &   0 &  -1 &   1 &   0 &   0   \\
   0 &   0 &   0 &  -1 &   0 &  -1   \\
\end{bmatrix}
\, ,
\end{align}
where the rows correspond to the nodes $v_0,v_1,v_2,v_3,v_4$, and the columns correspond to the arcs.
Since the affinity is absent on the twigs, the twisted root-to-node path matrix is the same as the standard one:
\begin{align}
\tilde{\bm S} = {\bm S} =
\begin{bmatrix}
   0 &   0 &   0 &   0  \\
   1 &   0 &   0 &   0  \\
   0 &   1 &   0 &   0  \\
   0 &   1 &   1 &   0  \\
   0 &   1 &   1 &   1  \\
\end{bmatrix}
\, ,
\end{align}
where the rows correspond to the nodes, and the columns correspond to the twigs $t_1,t_2,t_3,t_4$.

The edge currents are calculated as, for example, 
\begin{align}
j_{c_1} =& n_{\partial^+ c_1} - n_{\partial^- c_1} = (W_{v_2} - W_{v_1})/Z = 0 \, , \\
j_{c_2} =& n_{\partial^+ c_2} - \beta n_{\partial^- c_2} = (W_{v_2} - \beta W_{v_4})/Z 
\nonumber \\ =& 3 (1-\beta)/Z = \langle \! \langle w \rangle \! \rangle \, . 
\end{align}
Since the edge currents along $C_{c_1}$ vanish, whereas the edge currents along $C_{c_2}$ are identical, the cycle currents are given by, 
\begin{align}
{\bm i}^{\rm st} = \begin{bmatrix} i^{\rm st}_{C_{c_1}} \\ i^{\rm st}_{C_{c_2}} \end{bmatrix} = \begin{bmatrix} 0 \\ \langle \! \langle w \rangle \! \rangle \end{bmatrix}
\, . 
\end{align}
The edge traffic is given by, 
\begin{align}
g_{t_1} =& g_{t_2} = g_{c_1} = 2 \cdot 3 (2 \beta+1)/Z  \, , \\
g_{t_3} =& (\beta+1) 3^2/Z \, , \\
g_{t_4} =& (\beta+5) 3/Z \, , \\
g_{c_2} =& (5\beta+1) 3/Z \, . 
\end{align}
The edge traffic along $C_{c_1}$ is identical.

The remaining calculations are lengthy but can be done systematically. 
We relegate them to the Supplemental Material~\cite{SM}. 
The objective function (\ref{eqn:snr2}) is
\begin{align}
{\rm SNR}^2[{\bm f}] =& \frac{3 (8+7\beta) { f_{C_{c_1}} }^2}{2(1+2\beta)} \nonumber \\ &+ \frac{(8+7\beta)^3 { f_{C_{c_2}} }^2 }{46 + 214 \beta + 139 \beta^2 + 51 \beta^3}  \, ,
\end{align}
with the constraint $f_{C_{c_2}}=\langle \! \langle w \rangle \! \rangle$. 
The minimum is attained at $f_{C_{c_1}}=0$, which leads to (\ref{eqn:fano_2cycleG}).

The pseudo-entropy production rate is calculated as 
\begin{align}
\sigma_{\rm pseudo} = \langle \! \langle w \rangle \! \rangle \frac{2 (1-\beta)[23 (1+\beta^2)+62 \beta]}{3 (\beta+1) (5 \beta+1) (\beta+5) } \, . \label{eqn:pseudo_ent_2cycles} 
\end{align}

\section{Detailed calculations of Sec.~\ref{sec:LIBC}}
\label{app:LIBC}

\subsection{Node-state probability}

The node-state probability for a node $v_d = \partial^- t_d$ ($d=0,\cdots, \ell$, where $t_0=c$) on the cycle $C_c$, is computed as~\cite{Schnakenberg1976,Weidlich1978}, 
\begin{align}
n_{v_d}^{\rm st} =& W_d/Z \, , \label{eqn:n_cycle} 
\end{align}
where the total weight of directed spanning trees directed toward $v_{d}$ is calculated as, 
\begin{align}
W_d = \sum_{d'=d+1}^{\tilde{\ell}} \Gamma_{-t_{d+1}} \cdots  \Gamma_{-t_{d'-1}}  = \frac{\beta^{\tilde{\ell}-d}-1}{\beta-1}
\, . 
\end{align}
Here we exploited the fact that for an isolated subtree $T_d$ ($d=1,2,\cdots,\ell$), the product of backward transition rates along the twigs in the subtree directed toward $v_d$ is unity, $\prod_{b \in E(T_{d})} \Gamma_{-b}=1$.

In each subtree $T_d$, one can choose a path from the root node $v_d$ to a bottom leaf node $\tilde{v}_0 \in V(T_d)$ [dashed arcs in Fig.~\ref{fig:gTree_comp}]: 
\begin{align}
P_{\tilde{v}_0 \leftarrow v_d} = \left(\tilde{v}_0, b_{1}, \tilde{v}_{1}, b_{2}, \dots, \tilde{v}_{d-1}, b_{d}, {v}_d \right)  \, . 
\end{align}
For a node in the subtree $T_d$ at level $m=0,1,\cdots,d-1$, $\tilde{v}_{m} \in V(T_{d}) \backslash \{ v_d \}$, 
the total weight of directed spanning trees directed toward $\tilde{v}_{m}$ is given by $W_m^{(d)} =  \beta^{d-m} W_d$. 
The node-state probability is, 
\begin{align}
n_{\tilde{v}_m}^{\rm st \,(d)} =  W_m^{(d)}/Z \, , \label{eqn:n_subtree}
\end{align}
where the superscript $(d)$ specifies that the node $\tilde{v}_m$ is in the subtree $T_d$. 
Since at level $m$, there are 
\begin{align}
N_m^{(d)} =  (\alpha-1) \, \alpha^{d-1-m} \, , 
\quad
(m=0,1,\cdots,d-1) \, , 
\end{align}
such nodes, the normalization factor becomes
\begin{align}
Z = \sum_{d=1}^{\ell} \left( \sum_{m=0}^{d-1} N_m^{(d)}  W_m^{(d)} \right) + \sum_{d=0}^{\ell} W_{d} = \frac{\Delta}{(x-1)^2} \, , 
\end{align}
where $\Delta$ is defined in (\ref{eqn:aw}). 

\subsection{Edge current and traffic}

The steady-state edge current on a twig that lies on the cycle $t_d \in C_c$ ($d=1,2,\cdots,\ell$) is 
\begin{align}
j_{t_d}(n_v^{\rm st}) = n^{\rm st}_{v_{d-1}} - \beta n^{\rm st}_{v_d} = 1/Z \, ,
\end{align}
where we used unit forward rate and backward rate $\beta$, $\partial^+ t_d=v_{d-1}$, $\partial^- t_d=v_{d}$ and (\ref{eqn:n_cycle}). 
It is equal to the average reset current at the chord $c$, (\ref{eqn:aw}),
\begin{align}
\langle \! \langle w \rangle \! \rangle = j_{c}(n_v^{\rm st})=n^{\rm st}_{v_\ell}=1/Z \, .
\end{align}
The edge current on a twig of subtree $T_d$, $b_m \in E(T_d)$ is absent, 
\begin{align}
j_{b_m}^{(d)}(n_v^{\rm st}) = \beta n^{{\rm st} \, (d)}_{\tilde{v}_{m}} - n^{{\rm st} \, (d)}_{\tilde{v}_{m-1}} = 0 \, ,
\end{align}
where we used $\partial^+ b_m=\tilde{v}_m$, $\partial^- b_m=\tilde{v}_{m-1}$ and (\ref{eqn:n_subtree}). 
The result indicates that the steady-state current is purely supported on the cycle.

The traffic is calculated in the same manner. 
For a twig that lies on the cycle $t_d \in C_c$, the traffic is, 
\begin{align}
g_{t_d}(n_v^{\rm st}) = n^{\rm st}_{v_{d-1}} + \beta n^{\rm st}_{v_d} = \frac{2 \beta^{{\tilde{\ell}}-d+1} -\beta-1}{(\beta-1) Z} \, . 
\end{align}
The traffic on the chord is equal to the edge current, $g_{c}(n_v^{\rm st})= j_{c}(n_v^{\rm st})=1/Z$. 
For a twig in the subtree, $b_m \in E(T_d)$, the traffic is, 
\begin{align}
g_{b_m}^{(d)}(n_v^{\rm st}) = \beta n^{{\rm st} \, (d)}_{\tilde{v}_{m}} + n^{{\rm st} \, (d)}_{\tilde{v}_{m-1}}
= \frac{ 2 W_d \beta^{d-m+1} }{Z} \, . 
\end{align}

\subsection{Matrices and vectors}

We set the backward transition rate of the chord $c$ to a positive infinitesimal,
$\Gamma_{-c}=\eta$. 
Then the $(t_d,c)$ element of the matrix $\tilde{\bm F}$ is
\begin{align}
\tilde{F}_{t_d,c} = - \frac{\Gamma_{-c}}{\Gamma_{-t_d}} e^{A_c + A_{P_{v_\ell \leftarrow v_d }}}
= - \frac{\eta}{\beta} \frac{1}{\eta} \left( \frac{1}{\beta} \right)^{\ell-d} = - \beta^{d-{\tilde{\ell}}} \, ,
\end{align}
which is independent of $\eta$. 
From this expression, the Gram matrix (\ref{eqn:Gram_twist}) and the weighted cycle Laplacian (\ref{eqn:cycleLaplacian_bare}) become
\begin{align}
\tilde{\bm B} {\bm B}^{\mathsf{T}} =& \sum_{d=0}^{{\tilde{\ell}}-1} \beta^{-d} = \frac{1-\beta^{-{\tilde{\ell}}}}{1-\beta^{-1}} \, , \label{eqn:Gram_0} \\
\tilde{\bm B} {\bm G} \tilde{\bm B}^{\mathsf{T}}
=& \sum_{d=1}^{\ell} { \tilde{F}_{t_d,c} }^2 g_{t_d} + g_c = \frac{1}{Z} \left( \frac{1-\beta^{-{\tilde{\ell}}}}{1-\beta^{-1}} \right)^2
\, .
\end{align}

The elements of the vector $\mu_t$ ($t \in E(T)$) are calculated as follows. 
For a twig that lies on the cycle, $t_d \in C_c$, using the relation $V(T_{v_d}) = \bigcup_{d'=d}^{\ell} V(T_{d'})$ for $d=1,2,\dots,\ell$, we obtain
\begin{align}
\mu_{t_d} =& \sum_{v \in V \left ( T_{v_d} \right )} \frac{ e^{A_{P_{v \leftarrow v_d}}} }{\Gamma_{-t_d}} \nonumber \\
=& \sum_{d'=d}^{\ell} \frac{e^{A_{P_{v_{d'} \leftarrow v_d}}}}{\Gamma_{-t_d}} \left( 1 + \sum_{\tilde{v} \in V(T_{d'}) \setminus \{ v_{d'} \} } e^{A_{P_{\tilde{v} \leftarrow v_{d'}}}} \right) \, . 
\end{align}
The second term in parentheses is, 
\begin{align}
\sum_{m=0}^{d'-1} N_m^{(d')} e^{A_{P_{\tilde{v}_{m} \leftarrow v_{d'}}}} 
= \frac{ (x-\beta) (1-x^{d'})}{1-x} \, , 
\end{align}
where $x = \alpha \beta$. 
We therefore obtain, 
\begin{align}
\mu_{t_d} = \frac{1+\beta^d \left( \alpha^{\tilde{\ell}}-\beta^{-{\tilde{\ell}}} \right) -x^d }{x-1} \, . \label{eqn:mutd}
\end{align}

For a twig on the subtree, $b_m \in E(T_{d})$, 
\begin{align}
\mu_{b_m} = \sum_{\tilde{v} \in V \left ( T_{\tilde{v}_{m-1}} \right )} \frac{ e^{A_{P_{\tilde{v} \leftarrow \tilde{v}_{m-1}}}} }{\Gamma_{-b_m}} = \sum_{m'=0}^{m-1} x^{m'} = \frac{1-x^m}{1-x} \, .
\end{align}

By using (\ref{eqn:mutd}), 
\begin{align}
{\bm B} {\bm \mu} = \sum_{d=1}^{\ell} \mu_{t_d} = \frac{\beta ( 1 - \beta^{-{\tilde{\ell}}} ) (x^{{\tilde{\ell}}}-1) }{(\beta-1)(x -1)} - Z \, ,
\end{align}
which leads to, 
\begin{align}
{{\bm i}^{\rm st}}^{\sf{T}} {\bm B} {\bm \mu} = {{\bm i}^{\rm st}} \left( {\bm B} {\bm \mu} \right)^{\sf{T}}  = {\bm B} {\bm \mu} /Z
\, . \label{eqn:iBmu_long} 
\end{align} 
Then by substituting it and (\ref{eqn:Gram_0}) to (\ref{eqn:Gram_Phi}), we obtain 
\begin{align}
\tilde{\bm B} {\bm \Phi} {\bm B}^{\sf{T}} = \frac{\tilde{\ell}}{1- x^{\tilde{\ell}}} + \frac{x}{x-1} \, . \label{eqn:Gram_Phi_0}
\end{align} 

The weighted square norm of ${\bm \mu}$ (\ref{eqn:mu_sq_norm}) is, 
\begin{align}
{\bm \mu}^{\mathsf{T}} {\bm G} {\bm \mu} = \sum_{d=1}^{\ell} g_{t_d} {\mu_{t_d}}^2 + \sum_{d=1}^{\ell} \sum_{b \in E(T_{d})} g_{b} {\mu_{b}}^2 \, , \label{eqn:muGmu_long}
\end{align}
where in the second term of the right-hand side, 
\begin{align}
\sum_{b \in E(T_{d})} g_{b} {\mu_{b}}^2 = \sum_{m=0}^{d-1} N_m^{(d)} g_{b_{m+1}}^{(d)}{ \mu_{b_{m+1}} }^2 \, .
\end{align}
The vector in cycle space (\ref{eqn:BGmu}) is, 
\begin{align}
\tilde{\bm B} {\bm G} {\bm \mu} = - \sum_{d=1}^{\ell} \tilde{F}_{t_d,c} g_{t_d} \mu_{t_d} \, . \label{eqn:BGmu_long}
\end{align}
The explicit expressions for (\ref{eqn:muGmu_long}) and (\ref{eqn:BGmu_long}), which are lengthy, and further details of the calculations to derive (\ref{eqn:fano_ana}) are provided in the Supplemental Material~\cite{SM}.

\subsection{Steady-state TURs}
\label{app:ssTURs}

For the pseudo-entropy bound, 
\begin{align}
\frac{ \sigma_{\rm pseudo} }{ 2 \langle \! \langle w \rangle \! \rangle } = \sum_{d=1}^{\ell} \frac{j_{t_d}}{g_{t_d}} + \frac{j_{c}}{g_{c}} 
= \sum_{d=1}^{\ell} \frac{\beta-1}{2 \beta^{\tilde{\ell}-d+1}-\beta-1} + 1 \, . \label{eqn:pseudo}
\end{align}
In the limit $\ell \to \infty$, (\ref{eqn:pseudo}) diverges for $\beta<1$. 
For $\beta>1$, by replacing summation with integral, we obtain, 
\begin{align}
\frac{ \sigma_{\rm pseudo} }{ 2 \langle \! \langle w \rangle \! \rangle } \approx 1 + \frac{\beta-1}{(\beta+1) \ln \beta} \ln \frac{ 2 \beta^2}{(\beta-1) (1+2\beta)} \, . \label{eqn:pseudo_large} 
\end{align}
As $\beta \to 1^+$, the pseudo-entropy exhibits a logarithmic divergence,
$\sigma_{\rm pseudo}/( 2 \langle \! \langle w \rangle \! \rangle ) \approx 1 - \ln \sqrt{ 3(\beta-1)/{2} }$. 

For the mixed bound, 
\begin{align}
\frac{ \sigma_{\rm mix} }{2 \langle \! \langle w \rangle \! \rangle }
=&
\frac{1}{2} \sum_{d=1}^{\ell}  \ln \frac{\Gamma_{t_d} n_{v_{d-1}}}{\Gamma_{-{t_d}} n_{v_d}} + 1
\nonumber \\
=&
\frac{\ell}{2}  \ln \beta^{-1} + \frac{1}{2} \ln \frac{n_{v_{0}}}{n_{v_\ell}} + 1
\, , \label{eqn:mix}
\end{align}
where the first and second terms of the second line correspond to 
$\sigma_{\rm env}^{\rm bi}/(2 \langle \! \langle w \rangle \! \rangle)$
and 
$\sigma_{\rm sys}^{\rm bi}/(2 \langle \! \langle w \rangle \! \rangle)$
in (\ref{eqn:enbi}), respectively. 
The last term $1$ corresponds to the activity. 

In the long-computation-path limit, $\ell \to \infty$, 
$\sigma_{\rm tot}^{\rm bi}/\langle \! \langle w \rangle \! \rangle$ diverges as 
$-\ell \ln \beta$ for $\beta < 1$, whereas for $\beta > 1$ it approaches $\ln [\beta/(\beta-1)]$. 
This behavior leads to a nonanalyticity in the mixed bound 
$2 \langle \! \langle w \rangle \! \rangle/\sigma_{\rm mix}$ at $\beta=1$.

\bibliography{SSCFDUAL}

\end{document}